# Measurement of Proton Electromagnetic Form Factors in $e^+e^- \to p\bar{p}$ in the Energy Region 2.00–3.08 GeV

M. Ablikim,[1] M. N. Achasov,[10,d] P. Adlarson,[63] S. Ahmed,[15] M. Albrecht,[4] M. Alekseev,[62a,62c] A. Amoroso,[62a,62c] F. F. An,[1] Q. An,[59,47] Anita,[21] Y. Bai,[46] O. Bakina,[28] R. Baldini Ferroli,[23a] I. Balossino,[24a] Y. Ban,[37,l] K. Begzsuren,[26] J. V. Bennett,[5] N. Berger,[27] M. Bertani,[23a] D. Bettoni,[24a] F. Bianchi,[62a,62c] J. Biernat,[63] J. Bloms,[56] I. Boyko,[28] R. A. Briere,[5] H. Cai,[64] X. Cai,[1,47] A. Calcaterra,[23a] G. F. Cao,[1,51] N. Cao,[1,51] S. A. Cetin,[50b] J. Chai,[62c] J. F. Chang,[1,47] W. L. Chang,[1,51] G. Chelkov,[28,b,c] D. Y. Chen,[6] G. Chen,[1] H. S. Chen,[1,51] J. Chen,[16] M. L. Chen,[1,47] S. J. Chen,[35] X. R. Chen,[25] Y. B. Chen,[1,47] W. Cheng,[62c] G. Cibinetto,[24a] F. Cossio,[62c] X. F. Cui,[36] H. L. Dai,[1,47] J. P. Dai,[41,h] X. C. Dai,[1,51] A. Dbeyssi,[15] D. Dedovich,[28] Z. Y. Deng,[1] A. Denig,[27] I. Denysenko,[28] M. Destefanis,[62a,62c] F. De Mori,[62a,62c] Y. Ding,[33] C. Dong,[36] J. Dong,[1,47] L. Y. Dong,[1,51] M. Y. Dong,[1,47,51] S. X. Du,[67] J. Fang,[1,47] S. S. Fang,[1,51] Y. Fang,[1] R. Farinelli,[24a,24b] L. Fava,[62b,62c] F. Feldbauer,[4] G. Felici,[23a] C. Q. Feng,[59,47] M. Fritsch,[4] C. D. Fu,[1] Y. Fu,[1] Q. Gao,[1] Y. Gao,[60] Y. Gao,[59,47] Y. G. Gao,[6] B. Garillon,[27] I. Garzia,[24a,24b] E. M. Gersabeck,[54] A. Gilman,[55] K. Goetzen,[11] L. Gong,[36] W. X. Gong,[1,47] W. Gradl,[27] M. Greco,[62a,62c] L. M. Gu,[35] M. H. Gu,[1,47] S. Gu,[2] Y. T. Gu,[13] C. Y. Guan,[1,51] A. Q. Guo,[22] L. B. Guo,[34] R. P. Guo,[39] Y. P. Guo,[27] A. Guskov,[28] S. Han,[64] T. Z. Han,[9,j] X. Q. Hao,[16] F. A. Harris,[52] K. L. He,[1,51] F. H. Heinsius,[4] T. Held,[4] Y. K. Heng,[1,47,51] M. Himmelreich,[11,g] Y. R. Hou,[51] Z. L. Hou,[1] H. M. Hu,[1,51] J. F. Hu,[41,h] T. Hu,[1,47,51] Y. Hu,[1] G. S. Huang,[59,47] J. S. Huang,[16] L. Q. Huang,[60] X. T. Huang,[40] N. Huesken,[56] T. Hussain,[61] W. Ikegami Andersson,[63] W. Imoehl,[22] M. Irshad,[59,47] Q. Ji,[1] Q. P. Ji,[16] X. B. Ji,[1,51] X. L. Ji,[1,47] H. L. Jiang,[40] X. S. Jiang,[1,47,51] X. Y. Jiang,[36] J. B. Jiao,[40] Z. Jiao,[18] D. P. Jin,[1,47,51] S. Jin,[35] Y. Jin,[53] T. Johansson,[63] N. Kalantar-Nayestanaki,[30] X. S. Kang,[33] R. Kappert,[30] M. Kavatsyuk,[30] B. C. Ke,[42,1] I. K. Keshk,[4] A. Khoukaz,[56] P. Kiese,[27] R. Kiuchi,[1] R. Kliemt,[11] L. Koch,[29] O. B. Kolcu,[50b,f] B. Kopf,[4] M. Kuemmel,[4] M. Kuessner,[4] A. Kupsc,[63] M. Kurth,[1] M. G. Kurth,[1,51] W. Kühn,[29] J. S. Lange,[29] P. Larin,[15] L. Lavezzi,[62c] H. Leithoff,[27] T. Lenz,[27] C. Li,[38] Cheng Li,[59,47] D. M. Li,[67] F. Li,[1,47] G. Li,[1] H. B. Li,[1,51] H. J. Li,[9,j] J. C. Li,[1] J. W. Li,[44] Ke Li,[1] L. K. Li,[1] Lei Li,[3] P. L. Li,[59,47] P. R. Li,[31] W. D. Li,[1,51] W. G. Li,[1] X. H. Li,[59,47] X. L. Li,[40] X. N. Li,[1,47] Z. B. Li,[48] Z. Y. Li,[48] H. Liang,[59,47] H. Liang,[1,51] Y. F. Liang,[44] Y. T. Liang,[25] G. R. Liao,[12] L. Z. Liao,[1,51] J. Libby,[21] C. X. Lin,[48] D. X. Lin,[15] B. Liu,[41,h] B. J. Liu,[1] C. X. Liu,[1] D. Liu,[59,47] D. Y. Liu,[41,h] F. H. Liu,[43] Fang Liu,[1] Feng Liu,[6] H. B. Liu,[13] H. M. Liu,[1,51] Huanhuan Liu,[1] Huihui Liu,[17] J. B. Liu,[59,47] J. Y. Liu,[1,51] K. Liu,[1] K. Y. Liu,[33] Ke Liu,[6] L. Liu,[59,47] L. Y. Liu,[13] Q. Liu,[51] S. B. Liu,[59,47] T. Liu,[1,51] X. Liu,[31] X. Y. Liu,[1,51] Y. B. Liu,[36] Z. A. Liu,[1,47,51] Zhiqing Liu,[40] Y. F. Long,[37,l] X. C. Lou,[1,47,51] H. J. Lu,[18] J. D. Lu,[1,51] J. G. Lu,[1,47] X. L. Lu,[1] Y. Lu,[1] Y. P. Lu,[1,47] C. L. Luo,[34] M. X. Luo,[66] P. W. Luo,[48] T. Luo,[9,j] X. L. Luo,[1,47] S. Lusso,[62c] X. R. Lyu,[51] F. C. Ma,[33] H. L. Ma,[1] L. L. Ma,[40] M. M. Ma,[1,51] Q. M. Ma,[1] R. Q. Ma,[1,51] X. N. Ma,[36] X. X. Ma,[1,51] X. Y. Ma,[1,47] Y. M. Ma,[40] F. E. Maas,[15] M. Maggiora,[62a,62c] S. Maldaner,[27] S. Malde,[57] Q. A. Malik,[61] A. Mangoni,[23b] Y. J. Mao,[37,l] Z. P. Mao,[1] S. Marcello,[62a,62c] Z. X. Meng,[53] J. G. Messchendorp,[30] G. Mezzadri,[24a] J. Min,[1,47] T. J. Min,[35] R. E. Mitchell,[22] X. H. Mo,[1,47,51] Y. J. Mo,[6] C. Morales Morales,[15] N. Yu. Muchnoi,[10,d] H. Muramatsu,[55] A. Mustafa,[4] S. Nakhoul,[11,g] Y. Nefedov,[28] F. Nerling,[11,g] I. B. Nikolaev,[10,d] Z. Ning,[1,47] S. Nisar,[8,k] S. L. Niu,[1,47] S. L. Olsen,[51] Q. Ouyang,[1,47,51] S. Pacetti,[23b] Y. Pan,[59,47] M. Papenbrock,[63] A. Pathak,[1] P. Patteri,[23a] M. Pelizaeus,[4] H. P. Peng,[59,47] K. Peters,[11,g] J. Pettersson,[63] J. L. Ping,[34] R. G. Ping,[1,51] A. Pitka,[4] R. Poling,[55] V. Prasad,[59,47] H. Qi,[59,47] M. Qi,[35] T. Y. Qi,[2] S. Qian,[1,47] C. F. Qiao,[51] L. Q. Qin,[12] X. P. Qin,[13] X. S. Qin,[4] Z. H. Qin,[1,47] J. F. Qiu,[1] S. Q. Qu,[36] K. H. Rashid,[61,i] K. Ravindran,[21] C. F. Redmer,[27] M. Richter,[4] A. Rivetti,[62c] V. Rodin,[30] M. Rolo,[62c] G. Rong,[1,51] Ch. Rosner,[15] M. Rump,[56] A. Sarantsev,[28,e] M. Savrié,[24b] Y. Schelhaas,[27] K. Schoenning,[63] W. Shan,[19] X. Y. Shan,[59,47] M. Shao,[59,47] C. P. Shen,[2] P. X. Shen,[36] X. Y. Shen,[1,51] H. Y. Sheng,[1] H. C. Shi,[59,47] R. S. Shi,[1,51] X. Shi,[1,47] X. D. Shi,[59,47] J. J. Song,[40] Q. Q. Song,[59,47] Y. X. Song,[37,l] S. Sosio,[62a,62c] C. Sowa,[4] S. Spataro,[62a,62c] F. F. Sui,[40] G. X. Sun,[1] J. F. Sun,[16] L. Sun,[64] S. S. Sun,[1,51] T. Sun,[1,51] W. Y. Sun,[34] X. H. Sun,[1] Y. J. Sun,[59,47] Y. K. Sun,[59,47] Y. Z. Sun,[1] Z. J. Sun,[1,47] Z. T. Sun,[1] Y. T. Tan,[59,47] C. J. Tang,[44] G. Y. Tang,[1] X. Tang,[1] V. Thoren,[63] B. Tsednee,[26] I. Uman,[50d] B. Wang,[1] B. L. Wang,[51] C. W. Wang,[35] D. Y. Wang,[37,l] H. P. Wang,[1,51] K. Wang,[1,47] L. L. Wang,[1] L. S. Wang,[1] M. Wang,[40] M. Z. Wang,[37,l] Meng Wang,[1,51] P. L. Wang,[1] W. P. Wang,[59,47] X. Wang,[37,l] X. F. Wang,[31] X. L. Wang,[9,j] Y. Wang,[59,47] Y. Wang,[48] Y. F. Wang,[1,47,51] Z. Wang,[1,47] Z. G. Wang,[1,47] Z. Y. Wang,[1] Zongyuan Wang,[1,51] T. Weber,[4] D. H. Wei,[12] P. Weidenkaff,[27] H. W. Wen,[34] S. P. Wen,[1] U. Wiedner,[4] G. Wilkinson,[57] M. Wolke,[63] J. F. Wu,[1,51] L. H. Wu,[1] L. J. Wu,[1,51] Z. Wu,[1,47] L. Xia,[59,47] Y. Xia,[20] S. Y. Xiao,[1] Y. J. Xiao,[1,51] Z. J. Xiao,[34] Y. G. Xie,[1,47] Y. H. Xie,[6] T. Y. Xing,[1,51] X. A. Xiong,[1,51] Q. L. Xiu,[1,47] G. F. Xu,[1] J. J. Xu,[35] L. Xu,[1] Q. J. Xu,[14] W. Xu,[1,51] X. P. Xu,[45] F. Yan,[60] L. Yan,[62a,62c] W. B. Yan,[59,47] W. C. Yan,[2] Y. H. Yan,[20] H. J. Yang,[41,h] H. X. Yang,[1] L. Yang,[64] R. X. Yang,[59,47]

S. L. Yang,[1,51] Y. H. Yang,[35] Y. X. Yang,[12] Yifan Yang,[1,51] Z. Q. Yang,[20] Zhi Yang,[25] M. Ye,[1,47] M. H. Ye,[7] J. H. Yin,[1] Z. Y. You,[48] B. X. Yu,[1,47,51] C. X. Yu,[36] G. Yu,[1,51] J. S. Yu,[20] T. Yu,[60] C. Z. Yuan,[1,51] X. Q. Yuan,[37,l] Y. Yuan,[1] C. X. Yue,[32] A. Yuncu,[50b,a] A. A. Zafar,[61] Y. Zeng,[20] B. X. Zhang,[1] B. Y. Zhang,[1,47] C. C. Zhang,[1] D. H. Zhang,[1] H. H. Zhang,[48] H. Y. Zhang,[1,47] J. Zhang,[1,51] J. L. Zhang,[65] J. Q. Zhang,[4] J. W. Zhang,[1,47,51] J. Y. Zhang,[1] J. Z. Zhang,[1,51] J. Zhang,[1] J. Z. Zhang,[1,51] K. Zhang,[1,51] L. Zhang,[1] Lei Zhang,[35] S. F. Zhang,[35] T. J. Zhang,[41,h] X. Y. Zhang,[40] Y. Zhang,[59,47] Y. H. Zhang,[1,47] Y. T. Zhang,[59,47] Yang Zhang,[1] Yao Zhang,[1] Yi Zhang,[9,j] Yu Zhang,[51] Z. H. Zhang,[6] Z. P. Zhang,[59] Z. Y. Zhang,[64] G. Zhao,[1] J. Zhao,[32] J. W. Zhao,[1,47] J. Y. Zhao,[1,51] J. Z. Zhao,[1,47] Lei Zhao,[59,47] Ling Zhao,[1] M. G. Zhao,[36] Q. Zhao,[1] S. J. Zhao,[67] T. C. Zhao,[1] Y. B. Zhao,[1,47] Z. G. Zhao,[59,47] A. Zhemchugov,[28,b] B. Zheng,[60] J. P. Zheng,[1,47] Y. Zheng,[37,l] Y. H. Zheng,[51] B. Zhong,[34] C. Zhong,[60] L. Zhou,[1,47] L. P. Zhou,[1,51] Q. Zhou,[1,51] X. Zhou,[64] X. K. Zhou,[51] X. R. Zhou,[59,47] Xiaoyu Zhou,[20] Xu Zhou,[20] A. N. Zhu,[1,51] J. Zhu,[36] J. Zhu,[48] K. Zhu,[1] K. J. Zhu,[1,47,51] S. H. Zhu,[58] W. J. Zhu,[36] X. L. Zhu,[49] Y. C. Zhu,[59,47] Y. S. Zhu,[1,51] Z. A. Zhu,[1,51] J. Zhuang,[1,47] B. S. Zou,[1] and J. H. Zou[1]

(BESIII Collaboration)

[1]*Institute of High Energy Physics, Beijing 100049, People's Republic of China*
[2]*Beihang University, Beijing 100191, People's Republic of China*
[3]*Beijing Institute of Petrochemical Technology, Beijing 102617, People's Republic of China*
[4]*Bochum Ruhr-University, D-44780 Bochum, Germany*
[5]*Carnegie Mellon University, Pittsburgh, Pennsylvania 15213, USA*
[6]*Central China Normal University, Wuhan 430079, People's Republic of China*
[7]*China Center of Advanced Science and Technology, Beijing 100190, People's Republic of China*
[8]*COMSATS University Islamabad, Lahore Campus, Defence Road, Off Raiwind Road, 54000 Lahore, Pakistan*
[9]*Fudan University, Shanghai 200443, People's Republic of China*
[10]*G.I. Budker Institute of Nuclear Physics SB RAS (BINP), Novosibirsk 630090, Russia*
[11]*GSI Helmholtzcentre for Heavy Ion Research GmbH, D-64291 Darmstadt, Germany*
[12]*Guangxi Normal University, Guilin 541004, People's Republic of China*
[13]*Guangxi University, Nanning 530004, People's Republic of China*
[14]*Hangzhou Normal University, Hangzhou 310036, People's Republic of China*
[15]*Helmholtz Institute Mainz, Johann-Joachim-Becher-Weg 45, D-55099 Mainz, Germany*
[16]*Henan Normal University, Xinxiang 453007, People's Republic of China*
[17]*Henan University of Science and Technology, Luoyang 471003, People's Republic of China*
[18]*Huangshan College, Huangshan 245000, People's Republic of China*
[19]*Hunan Normal University, Changsha 410081, People's Republic of China*
[20]*Hunan University, Changsha 410082, People's Republic of China*
[21]*Indian Institute of Technology Madras, Chennai 600036, India*
[22]*Indiana University, Bloomington, Indiana 47405, USA*
[23a]*INFN Laboratori Nazionali di Frascati, I-00044 Frascati, Italy*
[23b]*INFN and University of Perugia, I-06100 Perugia, Italy*
[24a]*INFN Sezione di Ferrara, I-44122 Ferrara, Italy*
[24b]*University of Ferrara, I-44122 Ferrara, Italy*
[25]*Institute of Modern Physics, Lanzhou 730000, People's Republic of China*
[26]*Institute of Physics and Technology, Peace Avenue 54B, Ulaanbaatar 13330, Mongolia*
[27]*Johannes Gutenberg University of Mainz, Johann-Joachim-Becher-Weg 45, D-55099 Mainz, Germany*
[28]*Joint Institute for Nuclear Research, 141980 Dubna, Moscow region, Russia*
[29]*Justus-Liebig-Universitaet Giessen, II. Physikalisches Institut, Heinrich-Buff-Ring 16, D-35392 Giessen, Germany*
[30]*KVI-CART, University of Groningen, NL-9747 AA Groningen, Netherlands*
[31]*Lanzhou University, Lanzhou 730000, People's Republic of China*
[32]*Liaoning Normal University, Dalian 116029, People's Republic of China*
[33]*Liaoning University, Shenyang 110036, People's Republic of China*
[34]*Nanjing Normal University, Nanjing 210023, People's Republic of China*
[35]*Nanjing University, Nanjing 210093, People's Republic of China*
[36]*Nankai University, Tianjin 300071, People's Republic of China*
[37]*Peking University, Beijing 100871, People's Republic of China*
[38]*Qufu Normal University, Qufu 273165, People's Republic of China*
[39]*Shandong Normal University, Jinan 250014, People's Republic of China*
[40]*Shandong University, Jinan 250100, People's Republic of China*
[41]*Shanghai Jiao Tong University, Shanghai 200240, People's Republic of China*
[42]*Shanxi Normal University, Linfen 041004, People's Republic of China*

[43]Shanxi University, Taiyuan 030006, People's Republic of China
[44]Sichuan University, Chengdu 610064, People's Republic of China
[45]Soochow University, Suzhou 215006, People's Republic of China
[46]Southeast University, Nanjing 211100, People's Republic of China
[47]State Key Laboratory of Particle Detection and Electronics, Beijing 100049, Hefei 230026, People's Republic of China
[48]Sun Yat-Sen University, Guangzhou 510275, People's Republic of China
[49]Tsinghua University, Beijing 100084, People's Republic of China
[50a]Ankara University, 06100 Tandogan, Ankara, Turkey
[50b]Istanbul Bilgi University, 34060 Eyup, Istanbul, Turkey
[50c]Uludag University, 16059 Bursa, Turkey
[50d]Near East University, Nicosia, North Cyprus, Mersin 10, Turkey
[51]University of Chinese Academy of Sciences, Beijing 100049, People's Republic of China
[52]University of Hawaii, Honolulu, Hawaii 96822, USA
[53]University of Jinan, Jinan 250022, People's Republic of China
[54]University of Manchester, Oxford Road, Manchester M13 9PL, United Kingdom
[55]University of Minnesota, Minneapolis, Minnesota 55455, USA
[56]University of Muenster, Wilhelm-Klemm-Strasse 9, 48149 Muenster, Germany
[57]University of Oxford, Keble Road, Oxford OX13RH, United Kingdom
[58]University of Science and Technology Liaoning, Anshan 114051, People's Republic of China
[59]University of Science and Technology of China, Hefei 230026, People's Republic of China
[60]University of South China, Hengyang 421001, People's Republic of China
[61]University of the Punjab, Lahore-54590, Pakistan
[62a]University of Turin, I-10125 Turin, Italy
[62b]University of Eastern Piedmont, I-15121 Alessandria, Italy
[62c]INFN, I-10125 Turin, Italy
[63]Uppsala University, Box 516, SE-75120 Uppsala, Sweden
[64]Wuhan University, Wuhan 430072, People's Republic of China
[65]Xinyang Normal University, Xinyang 464000, People's Republic of China
[66]Zhejiang University, Hangzhou 310027, People's Republic of China
[67]Zhengzhou University, Zhengzhou 450001, People's Republic of China

The process of $e^+e^- \to p\bar{p}$ is studied at 22 center-of-mass energy points ($\sqrt{s}$) from 2.00 to 3.08 GeV, exploiting 688.5 pb$^{-1}$ of data collected with the BESIII detector operating at the BEPCII collider. The Born cross section ($\sigma_{p\bar{p}}$) of $e^+e^- \to p\bar{p}$ is measured with the energy-scan technique and it is found to be consistent with previously published data, but with much improved accuracy. In addition, the electromagnetic form-factor ratio ($|G_E/G_M|$) and the value of the effective ($|G_{\rm eff}|$), electric ($|G_E|$), and magnetic ($|G_M|$) form factors are measured by studying the helicity angle of the proton at 16 center-of-mass energy points. $|G_E/G_M|$ and $|G_M|$ are determined with high accuracy, providing uncertainties comparable to data in the spacelike region, and $|G_E|$ is measured for the first time. We reach unprecedented accuracy, and precision results in the timelike region provide information to improve our understanding of the proton inner structure and to test theoretical models which depend on nonperturbative quantum chromodynamics.

Despite the proton being one of the fundamental building blocks of atomic matter, its internal structure and dynamics are not well understood. Improving knowledge of these properties in terms of the proton's quark and gluonic degrees of freedom is one of the most challenging problems of modern nuclear physics. In addition, unsolved problems such as the proton-radius puzzle have recently attracted much attention [1].

The electric and magnetic form factors (FFs), $G_E(q^2)$ and $G_M(q^2)$, are fundamental quantities that can provide valuable insight into both the structure and dynamics of nucleons. FFs enter explicitly in the coupling of a virtual photon with the hadron electromagnetic current, and measurements can be directly compared to hadron models [1] giving, thereby, constraints in the description of the internal structure of hadrons. In the spacelike (SL) kinematic region (momentum transfer $q^2 < 0$), FFs have been studied in various electron-proton scattering experiments [Fig. 1(a)] since the 1950s, and are known with a precision of the order of a few percent. Over the past two decades several experiments have performed measurements that probe the timelike (TL) region ($q^2 > 0$), measured in annihilation reactions [Fig. 1(b)]. In most cases these measurements only extracted the effective FF ($G_{eff}$) or the ratio of $G_E$ and $G_M$ with uncertainties above 10%. Since the FFs in the SL and TL regions are connected via analyticity, precise knowledge of them in the TL region can help to solve problems in the SL region, such as the discrepancy found between the ratio $G_E/G_M$ determined via Rosenbluth separation and that found by experiments using polarized electron beams or targets [2].

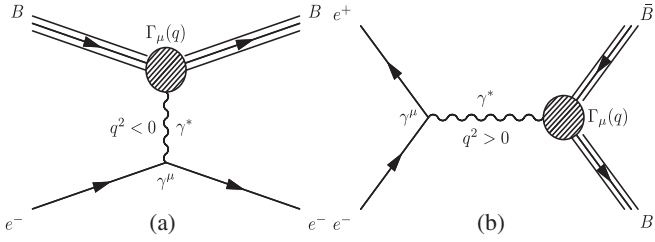

FIG. 1. Lowest-order Feynman diagrams for elastic electron-baryon scattering $e^-B \to e^-B$ (a), and for the annihilation process $e^-e^+ \to B\bar{B}$ (b). $B$ is a baryon.

The moduli of the FFs can be determined from the study of the angular distribution of the annihilation process [3], while the relative phase between the two FFs can be determined by measuring the polarization of the outgoing baryons.

The Born differential cross section as a function of the $e^+e^-$ center-of-mass (c.m.) energy squared $s$ reads [3]

$$\frac{d\sigma_{p\bar{p}}(s)}{d\Omega} = \frac{\alpha^2 \beta C}{4s} \Big[ |G_M(s)|^2 (1 + \cos^2\theta) + \frac{4m_p^2}{s} |G_E(s)|^2 \sin^2\theta \Big],$$  (1)

where $G_E$ and $G_M$ are the Sachs FFs, $\theta$ is the polar angle of the proton in the $e^+e^-$ c.m. frame, $m_p$ is the proton mass and $\beta = \sqrt{1 - 4m_p^2/s}$. The Coulomb enhancement factor, $C$, accounts for the electromagnetic interaction between the outgoing baryons. This factor is usually considered as a final-state interaction and it is $C = y/(1 - e^{-y})$ for point-like fermions with $y = \pi\alpha\sqrt{1 - \beta^2}/\beta$. Since the Coulomb interaction is long range, a pointlike correction is assumed when the two charged baryons are far apart.

In the TL region, the proton FFs can be accessed by three reactions: $e^+e^- \to p\bar{p}$ [4–10], $p\bar{p} \to e^+e^-$ [11–13], and the radiative-return process $e^+e^- \to p\bar{p}\gamma_{ISR}$ [14,15]. While there are many, generally consistent, measurements concerning the total $\sigma_{p\bar{p}}$, there are few and inconsistent data on the ratio $|G_E/G_M|$, mostly from PS170 [11] and

BABAR [14]. So far only two experiments [8,11] have been able to extract the value of $|G_M|$, which together with the knowledge of $|G_E/G_M|$ allows $|G_E|$ to be determined.

Precise measurements of FFs in the TL region may also be helpful for improved theoretical estimates of the proton radius [16,17]. From threshold energies to 3 GeV, the amplitude for the process is the sum of a leading term due to a bare formation process taking place on a time scale $1/\sqrt{q^2}$, and a relatively small perturbation associated with rescattering processes taking place on a longer time scale [16]. The combination of these effects is expected to lead to interesting phenomenology, in particular the superposition of small oscillations on an otherwise smooth dipole parameterization of the $G_{eff}$.

In this Letter, we present a study of the process $e^+e^- \to p\bar{p}$ at c.m. energies $\sqrt{s} = 2.00$–$3.08$ GeV, including a measurement of the Born cross section ($\sigma_{p\bar{p}}$), the electromagnetic FF ratio ($|G_E/G_M|$), the absolute value of the effective FF ($|G_{eff}|$) the magnetic FF ($|G_M|$) as well as, for the first time, the electric FF ($|G_E|$) of the proton using the energy-scan technique. The precision of our measurement is greatly improved with respect to that of previous experiments. Our results in the TL region have unprecedented precision with uncertainties comparable to FF measurements in the SL region.

The collision data were taken with the BESIII spectrometer at BEPCII. A detailed description of the detector and its performance can be found in Ref. [18]. The detector response, including the interaction of secondary particles with the detector material, is simulated using a GEANT4 [19] based program. Monte Carlo (MC) samples of 2.5 million $e^+e^- \to p\bar{p}$ events per energy point generated with CONEXC [20] are used for the efficiency determination and to calculate the correction factors for radiation up to next-to-leading order (NLO), as well as those for the vacuum polarization (VP). MC samples of QED background processes generated with BABAYAGA [21] and inclusive hadronic events generated with CONEXC [20] are used for background studies.

The final state of the process of interest is characterized by one proton and one antiproton. Hence, selected events must have exactly two charged tracks with opposite charge. A vertex fit is performed on both tracks under the hypothesis that the two particles in the final state are a proton and an antiproton to constrain them to one common vertex. A fit quality of $\chi^2 < 100$ is required to select candidate events. The opening angle between the proton and antiproton in the rest frame of the $e^+e^-$ c.m. system is required to be > 170° at 2.00 GeV and 2.05 GeV, > 175° at 2.1000 to 2.3094 GeV, and > 178° at 2.3864 to 3.0800 GeV. This condition ensures a back-to-back signature between the tracks. Cosmic-ray background is rejected by requiring $|T_{trk1} - T_{trk2}| < 4$ ns, where $T_{trk1}$ and $T_{trk2}$ are the measurements from the time-of-flight (TOF) system for each track. For $\sqrt{s}$ between 2.000 and 2.396 GeV, events are selected even if one of the two tracks does not hit the TOF system because of its low momentum. Finally, both tracks are required to be within an asymmetric momentum window around the average momentum, $p_{mean}$, determined from a fit to the momentum distribution after being boosted into the $e^+e^-$ c.m. system, namely $(p_{mean} - 4\sigma) < p < (p_{mean} + 3\sigma)$, where the spread $\sigma$ (standard deviation) is taken from the fit.

Particle identification (PID) is performed using the TOF and the $dE/dx$ measurement from the main drift chamber (MDC). At c.m. energies above 2.150 GeV this information is used to construct a probability for each track to conform to a particular (pion, kaon, electron, or proton) particle hypothesis to select the proton and antiproton candidates. For events at lower c.m. energies, the selection is made based on the normalized pulse height of the raw $dE/dx$ information. To remove Bhabha events, a requirement on $E/p$, defined as the ratio between the energy deposited by the track in the electromagnetic calorimeter (EMC) and its momentum measured in the MDC, is imposed for energy points above 2.150 GeV. Possible contamination from QED processes and hadronic final states are estimated to be less than 0.5% from studies performed on appropriate MC samples, and are neglected in the subsequent analysis.

With the number of events $N_{obs}$ selected, the cross section $\sigma_{p\bar{p}}$ of the process $e^+e^- \to p\bar{p}$ and $|G_{eff}|$ of the proton can be calculated with

$$\sigma_{p\bar{p}}(s) = \frac{N_{obs}}{\mathcal{L} \cdot \epsilon \cdot (1 + \delta)}, \qquad (2)$$

$$|G_{eff}(s)| = \sqrt{\frac{\sigma_{p\bar{p}}}{\frac{4\pi\alpha^2\beta C}{3s}\left(1 + \frac{2m_p^2}{s}\right)}}, \qquad (3)$$

where the efficiency $\epsilon$ and the correction factor $(1 + \delta) = \sigma_{obs}/\sigma_{Born}$ are determined with MC simulations. Here, $\sigma_{obs}$ is the cross section including NLO radiation and VP corrections, and $\sigma_{Born}$ is the born cross section. Results for the $\sigma_{p\bar{p}}$ and $G_{eff}$ measurement are summarized in Table I.

The FFs $|G_E|$ and $|G_M|$, or equivalently their ratio $|G_E/G_M|$ and $|G_M|$, can be determined from a fit to the proton angular distribution for energy points with a sufficiently high number of selected candidates. This is the case for 15 out of 22 energy points, as well as a combined sample of the individual data sets taken at c.m. energy points of 2.950, 2.981, 3.000, and 3.020 GeV with a luminosity weighted average energy of 2.988 GeV. The range of the angular analysis is limited to $\cos\theta$ from $-0.8$ to 0.8, because of the lack of efficiency in the gap between the barrel and end cap regions of the TOF system and EMC. The formula used to fit the proton angular distribution, deduced from Eqs. (1) and (2), can be expressed as

$$\frac{dN}{\epsilon(1 + \delta) \times d\cos\theta} = \frac{\mathcal{L}\pi\alpha^2\beta C}{2s} |G_M|^2 \Big[(1 + \cos^2\theta)$$
$$+ \frac{4m_p^2}{s}\left|\frac{G_E}{G_M}\right|^2 (1 - \cos^2\theta)\Big], \qquad (4)$$

TABLE I. The integrated luminosity, the number of $p\bar{p}$ events, the Born cross section $\sigma_{p\bar{p}}$, $|G_E/G_M|$, $|G_{eff}|$, $|G_E|$, and $|G_M|$.

| $\sqrt{s}$[GeV] | $\mathcal{L}$[pb$^{-1}$] | $N_{obs}$ | $\sigma_{p\bar{p}}$[pb] | $|G_{eff}|[10^{-2}]$ | $|G_E/G_M|$ | $|G_E|[10^{-2}]$ | $|G_M|[10^{-2}]$ |
|---|---|---|---|---|---|---|---|
| 2.0000 | 10.1 ± 0.1 | 5321 | 841.3 ± 11.5 ± 24.8 | 27.46 ± 0.19 ± 0.40 | 1.38 ± 0.10 ± 0.03 | 33.66 ± 1.23 ± 0.31 | 24.38 ± 0.99 ± 0.26 |
| 2.0500 | 3.34 ± 0.03 | 1703 | 753.4 ± 18.3 ± 23.5 | 24.94 ± 0.30 ± 0.39 | 1.24 ± 0.16 ± 0.04 | 29.10 ± 2.08 ± 0.40 | 23.48 ± 1.43 ± 0.42 |
| 2.1000 | 12.2 ± 0.1 | 5993 | 712.6 ± 9.2 ± 21.4 | 23.73 ± 0.15 ± 0.36 | 1.27 ± 0.09 ± 0.02 | 28.07 ± 1.10 ± 0.31 | 22.08 ± 0.74 ± 0.17 |
| 2.1250 | 108 ± 1 | 50312 | 660.0 ± 3.0 ± 19.7 | 22.69 ± 0.05 ± 0.34 | 1.18 ± 0.04 ± 0.01 | 25.62 ± 0.49 ± 0.18 | 21.65 ± 0.31 ± 0.13 |
| 2.1500 | 2.84 ± 0.02 | 1189 | 588.8 ± 17.1 ± 17.8 | 21.34 ± 0.31 ± 0.32 | 1.62 ± 0.24 ± 0.06 | 28.32 ± 1.89 ± 0.46 | 17.48 ± 1.51 ± 0.37 |
| 2.1750 | 10.6 ± 0.1 | 3762 | 491.0 ± 8.0 ± 14.8 | 19.44 ± 0.16 ± 0.29 | 1.19 ± 0.12 ± 0.02 | 22.08 ± 1.28 ± 0.28 | 18.55 ± 0.75 ± 0.16 |
| 2.2000 | 13.7 ± 0.1 | 4092 | 411.6 ± 6.4 ± 12.3 | 17.78 ± 0.14 ± 0.27 | 1.08 ± 0.10 ± 0.02 | 18.93 ± 1.20 ± 0.28 | 17.60 ± 0.63 ± 0.12 |
| 2.2324 | 14.5 ± 0.1 | 3644 | 341.9 ± 5.7 ± 10.1 | 16.21 ± 0.13 ± 0.24 | 0.85 ± 0.11 ± 0.03 | 14.48 ± 1.39 ± 0.42 | 16.98 ± 0.57 ± 0.17 |
| 2.3094 | 21.1 ± 0.1 | 2336 | 148.0 ± 3.1 ± 5.7 | 10.74 ± 0.11 ± 0.21 | 0.55 ± 0.16 ± 0.02 | 6.61 ± 1.72 ± 0.25 | 11.99 ± 0.44 ± 0.14 |
| 2.3864 | 22.5 ± 0.2 | 1851 | 122.0 ± 2.8 ± 3.6 | 9.87 ± 0.11 ± 0.15 | 0.54 ± 0.19 ± 0.02 | 5.98 ± 1.87 ± 0.19 | 10.99 ± 0.44 ± 0.07 |
| 2.3960 | 66.9 ± 0.5 | 5514 | 121.9 ± 1.6 ± 3.6 | 9.89 ± 0.07 ± 0.15 | 0.76 ± 0.10 ± 0.02 | 7.93 ± 0.86 ± 0.21 | 10.48 ± 0.27 ± 0.07 |
| 2.5000 | 1.10 ± 0.01 | 55 | 77.9 ± 10.5 ± 4.1 | 8.08 ± 0.55 ± 0.21 | ⋯ | ⋯ | ⋯ |
| 2.6444 | 33.7 ± 0.2 | 867 | 39.7 ± 1.3 ± 1.2 | 5.98 ± 0.10 ± 0.09 | 0.97 ± 0.24 ± 0.05 | 5.84 ± 1.13 ± 0.24 | 5.99 ± 0.37 ± 0.11 |
| 2.6464 | 34.0 ± 0.3 | 838 | 38.2 ± 1.3 ± 1.2 | 5.87 ± 0.10 ± 0.10 | 0.87 ± 0.27 ± 0.04 | 5.18 ± 1.30 ± 0.21 | 5.99 ± 0.37 ± 0.11 |
| 2.7000 | 1.03 ± 0.01 | 20 | 29.8 ± 6.7 ± 1.6 | 5.26 ± 0.59 ± 0.14 | ⋯ | ⋯ | ⋯ |
| 2.8000 | 4.76 ± 0.03 | 68 | 22.0 ± 2.7 ± 1.0 | 4.65 ± 0.28 ± 0.11 | ⋯ | ⋯ | ⋯ |
| 2.9000 | 105 ± 1 | 1010 | 15.0 ± 0.5 ± 0.5 | 3.95 ± 0.06 ± 0.06 | 0.54 ± 0.34 ± 0.03 | 2.31 ± 1.39 ± 0.11 | 4.29 ± 0.21 ± 0.06 |
| 2.9500 | 15.9 ± 0.1 | 118 | 11.7 ± 1.1 ± 0.4 | 3.53 ± 0.16 ± 0.07 | | | |
| 2.9810 | 16.1 ± 0.1 | 131 | 12.9 ± 1.1 ± 0.5 | 3.75 ± 0.16 ± 0.07 | 0.96 ± 0.39 ± 0.06 | 3.25 ± 1.09 ± 0.17 | 3.37 ± 0.28 ± 0.06 |
| 3.0000 | 15.9 ± 0.1 | 92 | 9.2 ± 1.0 ± 0.3 | 3.19 ± 0.17 ± 0.06 | | | |
| 3.0200 | 17.3 ± 0.1 | 97 | 9.0 ± 0.9 ± 0.3 | 3.16 ± 0.16 ± 0.05 | | | |
| 3.0800 | 157 ± 1 | 858 | 9.0 ± 0.3 ± 0.3 | 3.22 ± 0.05 ± 0.05 | 0.47 ± 0.45 ± 0.04 | 1.64 ± 1.53 ± 0.12 | 3.47 ± 0.18 ± 0.03 |

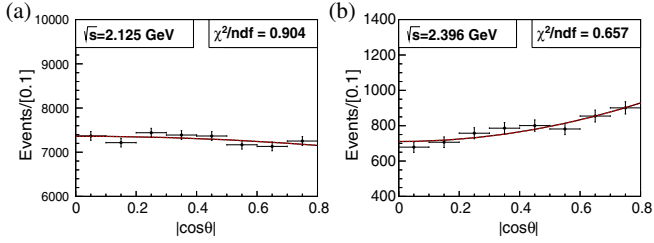

FIG. 2. Fit to the $|\cos\theta|$ distributions at (a) 2.125 GeV and (b) 2.396 GeV after the application of angular-dependent $\epsilon(1+\delta)$ factors.

where $\epsilon(\cos\theta)$ is the angular-dependent efficiency obtained from MC simulations. The correction factor, $(1+\delta)(\cos\theta)$, is calculated by dividing the $\cos\theta$ distribution of a MC sample generated with radiation up to NLO and VP corrections by the distribution of a sample generated with the Born process alone. A control sample of $e^+e^- \to p\bar{p}\pi^+\pi^-$ events is studied to determine correction factors for discrepancies between data and MC simulation in the angular-dependent efficiency.

After applying these corrections, the $|\cos\theta|$ distribution is fitted with Eq. (4). The results at 2.125 GeV and 2.396 GeV are shown in Fig. 2, while the results for all energy points are summarized in Table I. Fits to the $|\cos\theta|$ distributions as well as $\epsilon(1+\delta)(\cos\theta)$ distributions for all energy points can be found in the Supplemental Material [22].

The model used in the MC simulation takes as input $\sigma_{p\bar{p}}(s)$ and $|G_E(s)/G_M(s)|$. Therefore, the correction factors, and hence the measurements themselves, have a significant dependence on these inputs. For this reason, the complete analysis is performed in an iterative manner, where the obtained results are fed back into the MC simulation. After three iterations, the results for $\sigma_{p\bar{p}}(s)$ and $|G_E(s)/G_M(s)|$ are stable to within 1%.

Several sources of systematic uncertainties are considered in the determination of $\sigma_{p\bar{p}}$. The uncertainty associated with the knowledge of the reconstruction efficiency of the two charged tracks, as well as from the PID efficiency and the $E/p$ selection criteria, are studied with the $e^+e^- \to p\bar{p}\pi^+\pi^-$ control sample. The difference of the efficiency measured in data and MC simulation is assigned as the uncertainty, and it is found to be 1.0% for both tracking and PID, and 0.2% for the $E/p$ selection. The uncertainties due to the selection based on the TOF difference between the tracks, the angle between the tracks, and the momentum window are studied by varying the selection criteria. The uncertainty associated with the residual background contamination is estimated by comparing the populations of data and MC simulation in a momentum window of the same size as the signal region, but separated by $1\sigma$. The uncertainty from the luminosity measurement is found to be on a 1.0% level from Ref. [23]. The uncertainty due to the iterative MC-tuning procedure is assigned to be the difference between the nominal result and the result from the second-iteration step.

To assess the size of any bias from the choice of the used FF model in the MC simulations, we use the model from PHOKHARA [24] to generate an alternative set of MC events. The difference in the final result obtained with this new model and the default one is taken as the uncertainty.

Many of the uncertainties in the $|G_E/G_M|$ measurement are assigned with the same method as used in the cross-section analysis. This is true for all selection requirements, the uncertainties associated with the background, the iterative fit procedure, and the model used in the MC simulation. To account for any imperfections due to asymmetries between the fit model and the observed angular distributions, we fit the $\cos\theta$ distributions instead of the $|\cos\theta|$ ones and assign the difference as an uncertainty. The uncertainty from the luminosity measurement is taken as an independent systematic component for $|G_M|$, again taken from Ref. [23]. The total systematic uncertainties on $|G_E/G_M|$ range from 0.93% to 7.40%, while the total systematic uncertainties on $|G_M|$ range from 0.60% to 2.10%.

We study the energy dependence of $\sigma_{p\bar{p}}$ by fitting the expression

$$\sigma_{p\bar{p}}(s) = \begin{cases} \dfrac{e^{a_0}\pi^2\alpha^3}{s[1-e^{-\pi\alpha_s(s)/\beta(s)}][1+(\frac{\sqrt{s-2m_p}}{a_1})^{a_2}]}, & \sqrt{s} \leq 2.3094 \text{ GeV}, \\[4mm] \dfrac{2\pi\alpha^2\beta(s)C[2+(\frac{2m_p}{\sqrt{s}})^2]e^{a_3}}{3s^5[4\ln^2(\frac{\sqrt{s}}{a_4})+\pi^2]^2}, & \sqrt{s} > 2.3094 \text{ GeV}, \end{cases}$$

(5)

where $\alpha_s(s)$ is the strong coupling constant and $\alpha$ is the electromagnetic constant. The running coupling constant $\alpha_s(s)$ is parameterized as follows:

$$\alpha_s(s) = \left[\frac{1}{\alpha_s(m_Z^2)} + \frac{7}{4\pi}\ln\left(\frac{s}{m_Z^2}\right)\right]^{-1},$$

(6)

where $m_Z = 91.1876$ GeV is the mass of the $Z$ boson and $\alpha_s(m_Z^2) = 0.11856$ is the strong coupling constant at the $Z$ pole. Near the $p\bar{p}$ threshold, an alternative approach to the Coulomb enhancement factor should be considered in the cross section; concerning $B\bar{B}$, we have proposed gluon exchange. At large momentum transfer, the cross section is computed in perturbative QCD to leading order. Equation (5) takes into account strong-interaction effects near the threshold in a manner dependent on the perturbative-QCD prediction in the continuum region away from the threshold [16]. Correlations between the systematic uncertainties of the measurements at each energy point are taken into account. The results and meaning of the fit parameters are as follows: $a_0 = 0.80 \pm 0.08$ and $a_3 = 4.03^{+0.81}_{-0.47}$ are normalization constants, $a_1 = 0.35 \pm 0.01$ GeV is the QCD parameter near the threshold, $a_2 = 4.44 \pm 0.48$ is the $\sigma_{p\bar{p}}$ power-law dependence, which is related to the number of valence quarks, and $a_4 = 0.49^{+0.60}_{-0.37}$ GeV is the QCD parameter $\Lambda_{\text{QCD}}$ in the continuum region.

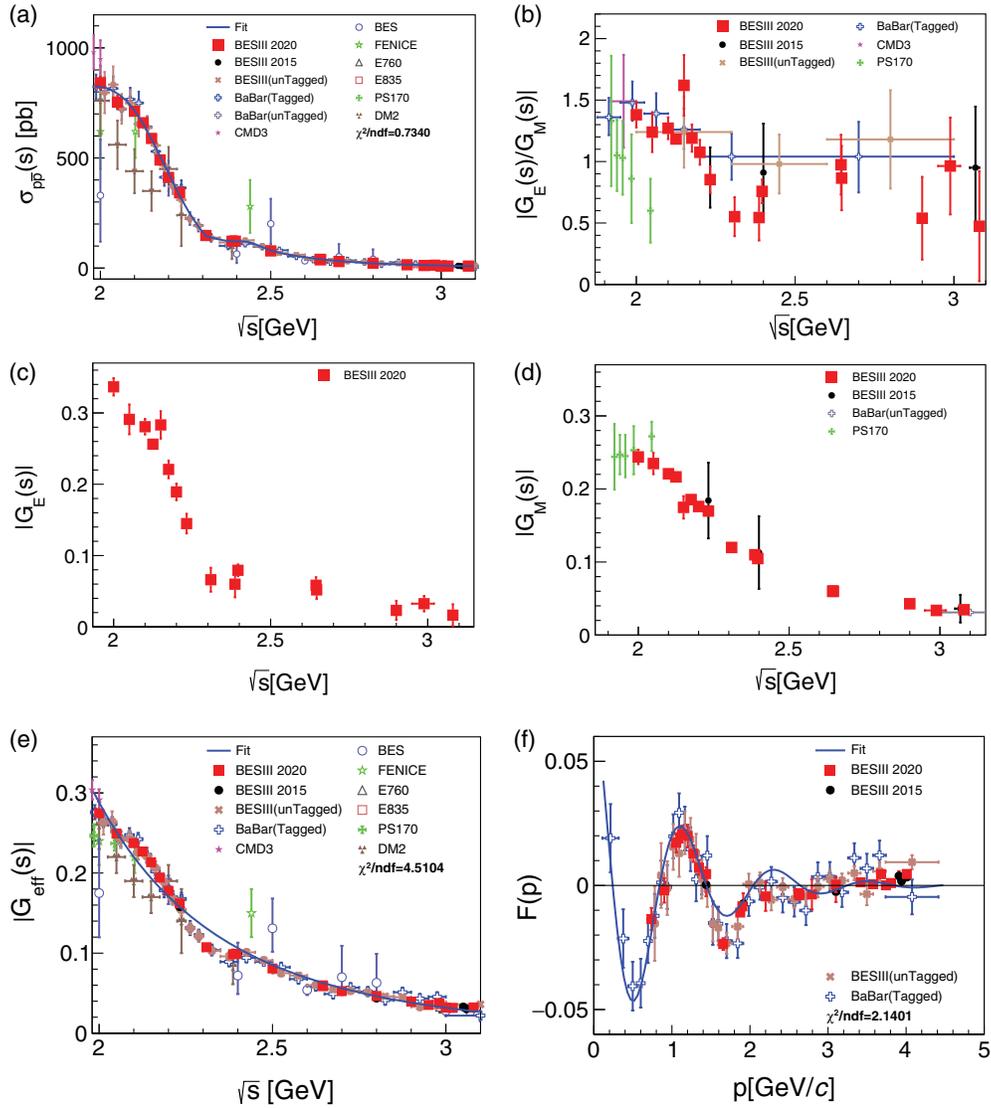

FIG. 3. Results from this analysis (red solid squares) including statistical and systematic uncertainties for (a) the $e^+e^- \to p\bar{p}$ cross section and a fit through the data (blue solid line); (b) the ratio $|G_E/G_M|$ of the proton; (c) the electric FF of the proton $|G_E|$; (d) the magnetic FF of the proton $|G_M|$; (e) the effective FF of the proton $|G_{eff}|$ and a fit through the data (blue solid line) by Eq. (7) suggested in Ref. [16]; (f) Proton effective FF values, after subtraction of the smooth function described by Eq. (7), as a function of the relative momentum $p$. Also shown are previously published measurements from BESIII [8,15], BABAR [14], CMD3 [10], BES [4], FENICE [9], E760 [12], E835 [13], PS170 [11], and DM2 [6]. $\chi^2 = \sum_i [f(x_i) - y_i]^2/\text{err}_i^2$, where $\text{err}_i$ is the error of the measured results including statistical and correlated systematic uncertainties, $f$ is the fit function, ndf is the number of degrees of freedom.

The data on the timelike $G_{eff}$ are best reproduced by the function proposed in Ref. [25],

$$|G_{eff}(s)| = \frac{\mathcal{A}}{(1 + \frac{s}{m_a^2})[1 - \frac{s}{0.71(\text{GeV}/c)^2}]^2}, \quad (7)$$

where $\mathcal{A} = 9.39 \pm 0.27$ and $m_a^2 = 7.72 \pm 0.54$ $(\text{GeV}/c)^2$ are obtained from our fit, illustrated in Fig. 3(e). The results indicate some oscillating structures which are clearly seen when the residuals are plotted as a function of the relative momentum $p$ of the $p\bar{p}$ pair [26]. The blue solid curve in Fig. 3(f) describes the periodic oscillations and has the form [26]

$$F_p = b_0^{osc} e^{-b_1^{osc} p} \cos(b_2^{osc} p + b_3^{osc}), \quad (8)$$

where $b_0^{osc} = 0.08 \pm 0.01$, $b_1^{osc} = 1.11 \pm 0.08$ $(\text{GeV}/c)^{-1}$, $b_2^{osc} = 5.23 \pm 0.13$ $(\text{GeV}/c)^{-1}$, and $b_3^{osc} = 0.31 \pm 0.17$ are obtained from our fit.

The data points and results of these fits are shown in Fig. 3 together with the data points for $|G_E/G_M|$, $|G_E|$, and $|G_M|$.

This Letter presents the most accurate measurement of the Born cross section of the process $e^+e^- \to p\bar{p}$, $\sigma_{p\bar{p}}$, for c.m. energies in the interval from 2.00–3.08 GeV. The uncertainties are dominated by systematics and range from 3.0% to 23.0%, respectively. Our data for $\sigma_{p\bar{p}}$ are found to be in good agreement with previously published results. The FF ratio $|G_E/G_M|$ is measured with total uncertainties around 10% for scan points ranging from low to intermediate energy. For the first time, the accuracy of the measured FF ratio in the TL region is comparable to that of data in the SL region. We have obtained an update of the FF measurement, especially for the ratio $|G_E/G_M|$, at c.m. energies of 2.2324 and 3.0800 GeV. We have tested the Coulomb enhancement factor hypothesis which depends on nonperturbative QCD. The oscillating structures in Refs. [15,26] are clearly seen in the $|G_{\text{eff}}|$ line shape.

Our measurement strongly favors the result of *BABAR* [14] over that of PS170 [11]. The magnetic form factor $|G_M|$ is measured for the first time over a wide range of energies with uncertainties of 1.6% to 3.9%, greatly improving the precision compared to previous measurements.

The BESIII Collaboration thanks the staff of BEPCII and the IHEP computing center for their strong support. This work is supported in part by National Key Basic Research Program of China under Contract No. 2015CB856700; National Natural Science Foundation of China (NSFC) under Contracts No. 11335008, No. 11375170, No. 11425524, No. 11475164, No. 11475169, No. 11605196, No. 11605198, No. 11625523, No. 11635010, No. 11705192, No. 11735014; the Chinese Academy of Sciences (CAS) Large-Scale Scientific Facility Program; the CAS Center for Excellence in Particle Physics (CCEPP); Joint Large-Scale Scientific Facility Funds of the NSFC and CAS under Contracts No. U1532102, No. U1532257, No. U1532258, No. U1732263, No. U1832103; CAS Key Research Program of Frontier Sciences under Contracts No. QYZDJ-SSW-SLH003, No. QYZDJ-SSW-SLH040; 100 Talents Program of CAS; INPAC and Shanghai Key Laboratory for Particle Physics and Cosmology; German Research Foundation DFG under Contract No. Collaborative Research Center CRC 1044, FOR 2359; Istituto Nazionale di Fisica Nucleare, Italy; Koninklijke Nederlandse Akademie van Wetenschappen (KNAW) under Contract No. 530-4CDP03; Ministry of Development of Turkey under Contract No. DPT2006K-120470; National Science and Technology fund; The Swedish Research Council; The Knut and Alice Wallenberg Foundation (Sweden); U.S. Department of Energy under Contracts No. DE-FG02-05ER41374, No. DE-SC-0010118, No. DE-SC-0010504, No. DE-SC-0012069; University of Groningen (RuG) and the Helmholtzzentrum fuer Schwerionenforschung GmbH (GSI), Darmstadt.

[a]Also at Bogazici University, 34342 Istanbul, Turkey.
[b]Also at the Moscow Institute of Physics and Technology, Moscow 141700, Russia.
[c]Also at the Functional Electronics Laboratory, Tomsk State University, Tomsk 634050, Russia.
[d]Also at the Novosibirsk State University, Novosibirsk 630090, Russia.
[e]Also at the NRC "Kurchatov Institute," PNPI, 188300 Gatchina, Russia.
[f]Also at Istanbul Arel University, 34295 Istanbul, Turkey.
[g]Also at Goethe University Frankfurt, 60323 Frankfurt am Main, Germany.
[h]Also at Key Laboratory for Particle Physics, Astrophysics and Cosmology, Ministry of Education; Shanghai Key Laboratory for Particle Physics and Cosmology; Institute of Nuclear and Particle Physics, Shanghai 200240, People's Republic of China.
[i]Also at Government College Women University, Sialkot—51310, Punjab, Pakistan.
[j]Also at Key Laboratory of Nuclear Physics and Ion-beam Application (MOE) and Institute of Modern Physics, Fudan University, Shanghai 200443, People's Republic of China.
[k]Also at Harvard University, Department of Physics, Cambridge, Massachusetts 02138, USA.
[l]Also at State Key Laboratory of Nuclear Physics and Technology, Peking University, Beijing 100871, People's Republic of China.